\documentclass[prl,
twocolumn,
 amsmath,amssymb,
 aps,
]{revtex4-2}

\usepackage{graphicx}
\usepackage{dcolumn}
\usepackage{bm}
\usepackage[utf8]{inputenc}
\usepackage{todonotes}
\usepackage{comment}
\usepackage{appendix}
\usepackage{soul}

\newcommand{\ri}{\mathrm{i}}

\begin{document}

\title{On the Relation of Exact Hydrodynamics to the Chapman--Enskog Series}

\author{Florian Kogelbauer}
\email{floriank@ethz.ch}\thanks{Corresponding Author}
\affiliation{Department of Mechanical and Process Engineering, ETH Z{u}rich, 8092 Z{u}rich, Switzerland}

\author{Ilya Karlin}
\email{ikarlin@ethz.ch}
\affiliation{Department of Mechanical and Process Engineering, ETH Z{u}rich, 8092 Z{u}rich, Switzerland}

\date{\today}

\begin{abstract}
We demonstrate that the Chapman--Enskog series is locally equivalent to the exact spectral closure defined on slow kinetic eigenmodes in the limit of vanishing Knudsen number. We further show that the Chapman--Enskog series diverges everywhere expect at the global equilibrium for an explicit example, while the exact spectrally closed hydrodynamics are defined globally for any Knudsen number.     
\end{abstract}

\maketitle

The Chapman--Enskog (CE) series is the fundamental and most widely used technique to derive a closed system of hydrodynamic equations from kinetic models, most notably the Boltzmann equation \cite{chapman1990mathematical}. Assuming that the kinetic evolution is close to local thermodynamic equilibrium, the CE series systematically expands the distribution function in powers of a small parameter — typically the Knudsen number Kn, which measures the ratio of the mean free path to a characteristic macroscopic length scale. One of the most prominent alternatives to the CE series is Grad's moment method, which relies on projection and subsequent truncation \cite{grad1949kinetic}. 

At leading order, the average fluxes obtained through the CE series give Newton's law of viscosity, Fourier's law of heat conduction and Fick's law of diffusion, thus reproducing the Navier--Stokes--Fourier equation \cite{ferziger1972matehematical} on a hydrodynamic level. This leading-order consistency of kinetic theory with continuum mechanics indicates the applicability of the CE series to fluid flow problems.
Despite its wide usage for the derivation of hydrodynamics, however, the convergence properties of the CE series are little discussed in kinetic theory. Regarding the CE series as an asymptotic Taylor expansion, the inclusion of higher-order terms should improve the quality of the obtained hydrodynamics, thus prompting the question about convergence of the CE series.

Notably, Grad \cite{grad1963asymptotic} has shown the asymptotic validity of the CE expansion, while McLennan \cite{mclennan1965convergence} shows that a convergent perturbation expansion of the linearized Boltzmann equation exists for small enough wave numbers. In some special cases, the CE series can even be summed explicitly, as exemplified for the three-component Grad system in \cite{gorban1996short}. We also refer to \cite{saint2014mathematical} for a mathematical treatment of the CE series with a focus on regularity and truncation. Let us remark that the convergence properties of the CE series are of interest in the context of relativistic kinetic theory \cite{denicol2016divergence} and in granular media \cite{santos2008does} as well. 

While the CE series recovers the classical equations of fluid dynamics for small Knudsen numbers, linear higher-order hydrodynamics such as the Burnett equation \cite{burnett1936distribution} exhibit nonphysical instabilities due to a sign-change of the dissipation relation in frequency space  \cite{bobylev1982chapman}, called Bobylev instability, thus questioning the global validity of the CE series. Indeed, Santos \cite{santos1986divergence} shows the divergence of the CE series for shear flows from the BGK model. In this note, we aim to clarify the validity and convergence of the CE series in light of the interpretation of hydrodynamics as a slow manifold. In particular, we show how the CE is intimately linked to the spectral theory of linearized kinetic operators. 


Before we continue with the analysis of the CE series from a geometric perspective, let us recall the fundamental question of kinetic theory: How can we  derive hydrodynamic equations from the evolution of distribution functions? Historically, the ingenious ideas of Maxwell \cite{maxwell1867iv} and Boltzmann \cite{boltzmann1872weitere} that lead to the modern development of kinetic theory consist in the abstraction evolution of a distribution function from which the evolution of physical observables such as the temperature or the velocity field of a gas can be inferred by taking expectations. The moment closure problem \cite{grad1949kinetic} arises when we try to derive a self-consistent evolution of the macroscopic observables from a kinetic description. Any closure assumption is thus equivalent to a constitutive law that expresses higher-order fluxes in terms of lower-order observables. From a mathematical point of view, the relation between the kinetic and the continuum description of fluids and the moment closure problem derived from it lies at the heart of Hilbert's sixth problem \cite{hilbert1989grundzuge}.

In this work, we contrast the CE series to a different viewpoint by postulating that hydrodynamics correspond to kinetic dynamics restricted to a special subset of distribution functions \cite{gorban2005invariant}. These geometric ideas are formalized in the theory of slow manifolds for kinetic equations \cite{kogelbauer2024rigorous} by assuming that hydrodynamics correspond to an invariant manifold in the phase space of a kinetic equation. We mention in passing that the theory of slow manifolds itself originated in atmospheric sciences \cite{lorenz1986existence} and has been applied to a multitude of physical problems \cite{mackay2004slow}, while its relation to spectral properties has recently been formalized mathematically \cite{cabre2003parameterization}. 

In the following, we will connect the convergence properties of the CE series with the exact, optimal hydrodynamics given by the slow spectral closure. To this end, we consider the Boltzmann equation for a rarefied gas, 
\begin{equation}\label{maineq}
    \frac{\partial f}{\partial t} + \bm{v}\cdot\nabla f = \frac{1}{\rm Kn}Q(f,f),
\end{equation}
where $f$ is the distribution function, Kn is Knudsen number and $Q$ is the binary collision term with general cross section \cite{chapman1990mathematical}.
To illustrate the relation between hydrodynamics and kinetic theory, we will focus on ensembles close to global equilibrium by considering the linearized Boltzmann equation, as widely used in spectroscopy \cite{wang2019bulk} and classical problems in rarefied gas flows, such as the temperature jump or the thermal slip \cite{siewert2003viscous} . 

For linear kinetic equations, the time evolution of the distribution function is equivalent to the spectral analysis of the linearized kinetic operator. Analogous to energy states of the Schr\"{o}dinger equation in quantum physics, we are interested in kinetic eigenstates of the linearized Boltzmann equation. In our interpretation, hydrodynamics then correspond to special parts of the spectrum and their eigenstates that are separated from the remainder of the spectrum by the magnitude of their real part \cite{kogelbauer2024rigorous}.

To formalize the reduction onto slow eigenmodes, let $L_Q$ denote the collision kernel linearized around the global Maxwellian equilibrium. The operator $L_Q$ is self-adjoint, positive semi-definite and allows for countably many isolated eigenvalues of finite multiplicity \cite{cercignani1988boltzmann}. Its center directions are called collision invariants and correspond to the global conservation of mass, momentum and energy.

The hydrodynamics as dynamics on a slow manifold are thus encoded in the eigenvalues $\lambda$ of the linearized kinetic operator,
\begin{equation}\label{eigenvalues}
    -\ri (\bm{k}\cdot v)f + \frac{1}{\rm Kn}L_Q(f) = \lambda(\bm{k},{\rm Kn}) f,
\end{equation}
where $\bm{k}$ denotes the spatial wave vector and the left-hand side of \eqref{eigenvalues} is the linearized Boltzmann equation in frequency space. Because of rotational symmetries of the collision operator $L_Q$ and scaling properties of eigenstates \cite{kogelbauer2024rigorous}, the Knudsen number and the wave number  $k = |\bm{k}|$ are coupled in the eigenvalues in \eqref{eigenvalues} as
\begin{equation}\label{coupling}
    \lambda(\bm{k},{\rm Kn}) = \frac{1}{\rm{Kn}} \hat{\lambda}(k\times  \rm{Kn}). 
\end{equation}

\begin{figure}
    \centering
    \includegraphics[width=0.8\linewidth]{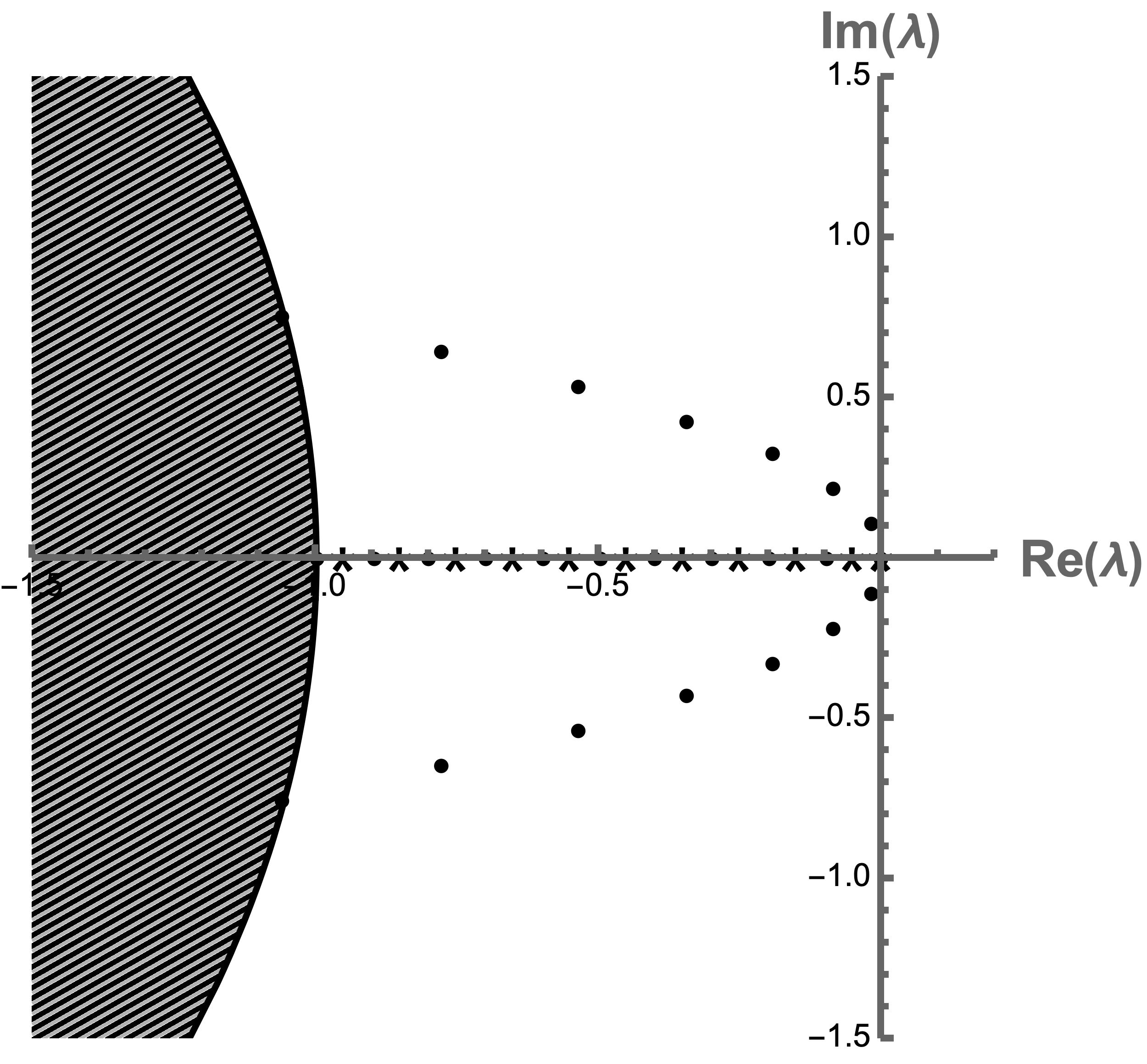}
    \caption{Schematic depiction of the typical operator spectrum of a linear kinetic operator: The essential spectrum (striped region) corresponds to fluctuations, while the branches of the point spectrum, parametrized by spatial wave number, correspond to slow modes and define hydrodynamics. The five hydrodynamic modes bifurcate out of the collision invariants and merge with the essential spectrum at a specific wave number (criticality). }
\label{Fig_spectral}
\end{figure}

As pointed out, the understanding of spectral properties of the linearized Boltzmann operator is crucial for the slow manifold reduction. Historically, along with the fundamental developments in quantum physics, the spectral properties of the linearized Boltzmann operator were actually among the first to be studied systematically by Hilbert \cite{hilbert1989grundzuge}. The fundamental spectral properties of the linear Boltzmann operator, including a perturbative analysis of hydrodynamic branches of eigenvalues, were worked out in  \cite{ellis1975first}: There exist five branches of hydrodynamic eigenvalues in \eqref{eigenvalues} - the diffusion mode, the double degenerate shear mode and a complex conjugated pair of acoustic modes - bifurcating from the five-fold degenerate eigenvalue zero for increasing wave number. Each eigenvalue branch $\lambda_j(k)$ exists up to a critical wave number $k_{\rm crit, j}$ at which it merges with the essential spectrum. A typical operator spectrum of a linear kinetic equation is shown in Figure \ref{Fig_spectral}. The eigenvalue branches can be calculated explicitly for certain models, such as the BGK operator \cite{kogelbauer2024exact}. 

The properties of eigenvalue branches \eqref{eigenvalues} lie at the heart of the spectral closure \cite{kogelbauer2024rigorous,kogelbauer2025exact}. The most prominent feature of the spectrally closed hydrodynamics is the emergence of a hidden length scale through the critical wave number. On an operator spectrum level, this is equivalent to a finite existence range of wave numbers for the hydrodynamic modes \cite{nishida1976global}. In terms of constitutive laws, the slow spectral closure can be interpreted as the dynamically optimal closure assumption \cite{kogelbauer2025learning}. 

Let us turn to the main question of this Letter: What is the relation of the CE series to the hydrodynamic eigenvalues \eqref{coupling}? The fundamental observation is that the CE series for linear kinetic models approximates the eigenvalue function \eqref{eigenvalues} in Knudsen number and, thanks to the coupling \eqref{coupling}, defines an expansion in wave number as well. Due to criticality in wave number, however, this approximation of the CE cannot be a global representation of the eigenvalue function.

Indeed, despite the five-fold degeneracy of the zero eigenvalue for vanishing wave number, the eigenvalues branches are locally analytic functions in wave number \cite{ellis1975first,mclennan1965convergence} and hence, thanks again to the coupling \eqref{coupling}, the eigenvalue branches are locally analytic functions of Knudsen number as well. From analytical spectral perturbation theory \cite{kato2013perturbation} as a composition of analytic functions guarantees that the slow spectral closure is equivalent to the CE series to all orders for small enough Knudsen number. The spectrally closed hydrodynamics at each wave number are thus equivalent to the corresponding terms of the CE series order by order.

While the CE series is locally equivalent to the spectrally closed hydrodynamics, globally the exact hydrodynamics deviate from the CE series due to criticality. Indeed, we will give an example of a kinetic model for which the CE series is divergent for every Knudsen number except for ${\rm Kn} = 0$, while the spectrally closed hydrodynamics define a unique optimal reduction for any Knudsen number. 

Let us comment on the relation to truncation in the CE series and how they translate to the spectral closure. The Bobylev instability arises due to the polynomial approximation of the CE series to the eigenvalue function $\tau k\mapsto \hat{\lambda}(\tau k)$ 
of the Boltzmann equation, which in itself is compactly supported and remains sign-definite. Criticality, i.e., the existence of a hydrodynamic mode for a finite range of wave numbers, thus implies that the true spectral hydrodynamics can never be globally represented by an analytic function for kinetic models. This feature poses a mathematical obstruction, to explain why the CE series cannot represent the eigenvalue branch globally: A compactly supported function cannot be globally represented by a Taylor series \cite{rudin1987real}. In particular, this holds for the full Boltzmann equation, which exhibits criticality \cite{dudynski2013spectral}. 

We mention, however, that certain CE series obtained from moment truncations, which do not exhibit criticality due to projection onto finitely many moments, can converge. In fact, the exact sum of the CE series performed in \cite{karlin2014non} for the three-component Grad system is equivalent to the spectral analysis carried out in \cite{kogelbauer2020slow}, showing that the eigenvalue branches do not exhibit criticality and are, in fact, globally analytic functions of wave number. This explains why certain CE series converge, while others drastically diverge. We stress again that the absence of criticality in Grad-type models is a feature of the moment truncations, while the full Boltzmann equation exhibits criticality. 

To illustrate the relation of the CE series with the slow spectral closure more explicitly, we consider the following one-dimensional linear kinetic equation,
\begin{equation}\label{kin1d}
    \frac{\partial f}{\partial t} + v\frac{\partial f}{\partial x} = -\frac{1}{\tau}\left(f-\frac{\rho[f]}{\sqrt{2\pi}}e^{-\frac{v^2}{2}}\right),
\end{equation}
for the distribution function $f$, the relaxation time $\tau$, which plays the role of the Knudsen number in the following, and the mass density
\begin{equation}
    \rho[f](x,t) = \int_{-\infty}^\infty f(v,x,t)dv.
\end{equation}
Equation \eqref{kin1d} appears in shear-mode reductions \cite{cercignani1963flow} of the three-dimensional BGK equation \cite{bhatnagar1954model} and is arguably the simplest nontrivial kinetic model to illustrate our considerations. 

The essential spectrum \cite{kato2013perturbation}, corresponding to fluctuations, is given by the vertical line $\{\Re \lambda = -1/\tau\}$, while the point spectrum consists of a single eigenvalue at each wave number $k$, the diffusion mode $\lambda_d(k)$. The spectrally closed dynamics for the Fourier transform $\hat{\rho}(k)$ of the density were derived in \cite{kogelbauer2021non} are simplify given by 
\begin{equation}
    \frac{\partial\hat{\rho}}{\partial t}(k) = \lambda_d(k,\tau )\hat{\rho}(k), 
\end{equation}
where the slow diffusion mode is given as the solution to the transcendental equation
\begin{equation}\label{deflambda}
    Z\left(\ri\frac{\tau \lambda_d(k,\tau)+1}{\tau k}\right) = \ri \tau k,
\end{equation}
for the plasma dispersion function \cite{fried2015plasma}:
\begin{equation}
    Z(\zeta) = \frac{1}{\sqrt{2\pi}}\int_{-\infty}^\infty dv \frac{e^{-\frac{v^2}{2}}}{v-\zeta}. 
\end{equation}

Because of asymptotic degeneracies in the dependence of $k$ and $\lambda_d$ in \eqref{deflambda}, the Lagrange Inversion Theorem \cite{abramowitz1948handbook} cannot be applied to \eqref{deflambda} directly and no closed-form solution for $\lambda_d$ in terms of elementary functions can be given.
The critical wave number for the eigenvalue branch can be derived explicitly and is given by 
\begin{equation}
    k_{\rm crit} = \sqrt{\frac{\pi}{2}}\frac{1}{\tau},
\end{equation}
beyond which \eqref{deflambda} does not admit a solution. We further note that the singularity of $Z$ at the critical wave number is essential and corresponds to a branch cut of the two sheets of $Z$, each admitting an extension to the whole complex plane. 

\begin{figure}
    \centering
    \includegraphics[width=0.9\linewidth]{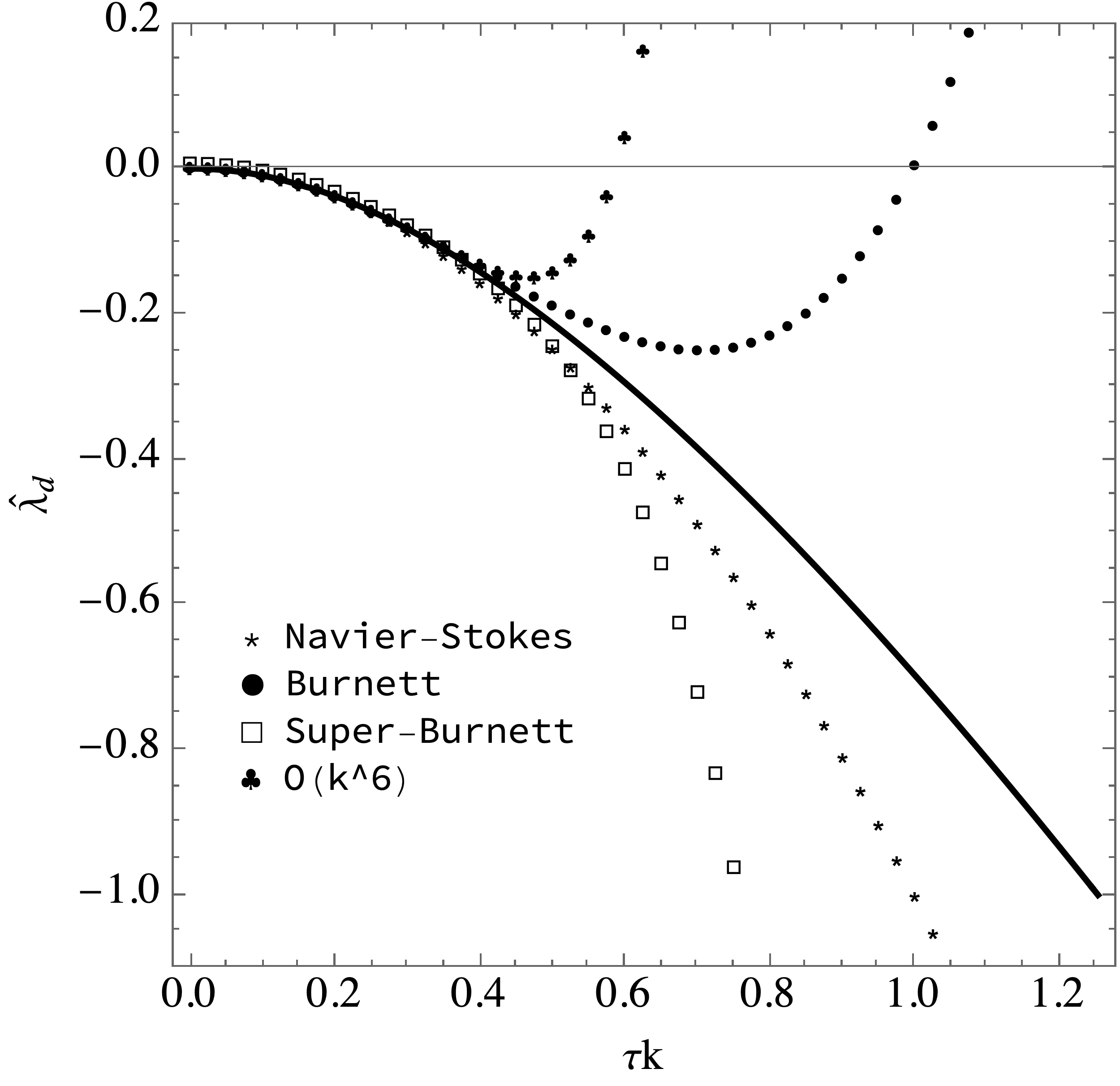}
    \caption{Comparison of the shear mode (solid black line) to CE approximations for the one-dimensional kinetic model: Higher-order terms in the CE series provide better approximations at the origin, but deviate further for larger wave numbers. Approximations at odd powers of $k^2$ change sign and develop Bobylev-type instabilities.}
\label{Fig_comp_shear_CE}
\end{figure}

Expanding equation \eqref{deflambda} in powers of wave number gives the series 
\begin{small}
\begin{equation}\label{expandlambda}
 \hat{\lambda}_d(\tau k) = -(\tau k)^2 + (\tau k)^4 - 4(\tau k)^6 + 27(\tau k)^8 + \mathcal{O}((\tau k)^{10}),
\end{equation}
\end{small}
whose leading-order truncation corresponds to the one-dimensional linear analogue of the Navier--Stokes dynamics, which just reduces to the heat equation, while its higher-order truncations correspond to the one-dimensional linear Burnett- and super-Burnett-type equations \cite{garcia2008beyond}. Figure \ref{Fig_comp_shear_CE} compares the successive approximations in wave number, i.e., the terms of the CE series, to the full $\hat{\lambda}_d$-function given as solution to \eqref{deflambda}. 
 
We remark that it has been argued that the reduced wave number at which the transport coefficients of the Burnett equation change sign can be interpreted as a critical wave number beyond which Burnett equation loses its validity \cite{uribe2000bobylev}. Figure \ref{Fig_comp_shear_CE} indicates that this sign-change wave number occurs before the actual critical wave number and decreases with Knudsen number. It is thus not actually related to a loss of validity of the CE series, but much rather a feature of the polynomial approximation of the true eigenvalue function. 

Let us take a closer look at the expansion of the diffusion mode in wave number. The first few CE coefficients are given by  and coincide with series $A000699$ in Sloane's Online Encyclopedia of Integer Sequences \cite{oeisA000699}. Sequence $A000699$ has a combinatorial interpretation as the number of irreducible chord diagrams with $2n$ nodes \cite{flajolet2000analytic} and is indeed equivalent to the Taylor expansion in wave number of the solution to \eqref{deflambda}, see \cite{marie2012chord}. As shown in \cite{borinsky2018generating}, the  $A000699$ sequence satisfies the asymptotics,
\begin{equation}
    \lambda_n \sim (2n-1)!!,
\end{equation}
thus proving that the CE series of equation \eqref{kin1d} is factorially divergent for each non-zero Knudsen number. At the origin, the coefficients trivially sum up to zero, showing that the CE series is strongly divergent everywhere except at one point. On the other hand, the derivatives of the diffusion mode coincide with the entries of the CE series around the origin to any order, as proved before. 

Super-exponential growth of Taylor coefficients is frequently encountered in quantum field theory, see e.g. \cite{bender1969anharmonic} for the factorial growth of the coupling constant coefficients of the partition function in zero‐dimensional $\phi^4$ theory. This is not a coincidence. Indeed, we note that the asymptotic expansion \eqref{expandlambda} appeared in \cite{kleitman1970proportions} in the context of a spinor model and that an iterative scheme to obtain the CE coefficients based on the Schwinger--Dyson series was presented in \cite{karlin2014non}.

The appearance of a strongly divergent asymptotic expansion in the context of kinetic theory is naturally connected to renormalization theory \cite{glimm2012quantum}. The idea to regularize the Chapman--Enskog series dates at least back to Rosenau \cite{rosenau1989extending} who proposed a phenomenological non-local closure relation, which can be interpreted as an analogue of a renormalization series. We also mention \cite{karlin2007renormalization} in the context of renormalization group techniques applied to kinetic theory. It would be interesting to exploit the connection between the one-dimensional kinetic model \eqref{kin1d} along with its CE series \eqref{expandlambda} to the spinor model \cite{kleitman1970proportions} and its renormalization group in more detail. 

We conclude with a summary of the results. We demonstrated that the theory of slow spectral closure is equivalent to the CE series to all orders in the limit of vanishing Knudsen number. 
While the slow spectral closure gives a complete description of the dynamically optimal reduced dynamics for any Knudsen number, higher-order contributions of the CE might lead to artifacts such as Bobylev's instability. We illustrated the relation of the exact spectral closure to the CE series for an explicit example, where the spectral closure gives a closed-form expression for the slow diffusion-type eigenvalue function, while the CE series is divergent everywhere expect for ${\rm Kn} = 0$.

The connection of the CE series to quantum field theories, especially the spinor model in \cite{kleitman1970proportions}, offers an intriguing new perspective: Can the theory of slow spectral closure be equivalently applied to spinor models? If so, what is the relation of the exact, spectrally closed system to dynamics obtained from the application of renormalization group techniques? \\

\begin{acknowledgments}
This work was supported by European Research Council (ERC) Advanced Grant  834763-PonD and
Swiss National Science Foundation (SNSF) grant No. 200021-228065.
Computational resources at the Swiss National  Super  Computing  Center  CSCS  were  provided  under the grant s1286.
\end{acknowledgments}

\bibliographystyle{apsrev4-1}
\bibliography{Spectral_CE}

\end{document}